\documentclass[aps,prl,twocolumn,showpacs]{revtex4}

\usepackage{graphicx}

\begin{document}

\title{Exact results for the Barab\'asi model of human dynamics}

\author{Alexei V\'azquez}

\affiliation{Department of Physics and Center for Complex Networks 
Research, University of Notre Dame, Notre Dame, IN 46556, USA}

\begin{abstract}

Human activity patterns display a bursty dynamics, with interevent times
following a heavy tailed distribution.  This behavior has been recently
shown to be rooted in the fact that humans assign their active tasks
different priorities, a process that can be modeled as a priority queueing
system [A.-L. Barab\'asi, Nature {\bf 435}, 207 (2005)]. In this work we
obtain exact results for the Barab\'asi model with two tasks, calculating
the priority and waiting time distribution of active tasks. We demonstrate
that the model has a singular behavior in the extremal dynamics limit,
when the highest priority task is selected first. We find that
independently of the selection protocol, the average waiting time is
smaller or equal to the number of active tasks, and discuss the asymptotic
behavior of the waiting time distribution. These results have important
implications for understanding complex systems with extremal dynamics.

\end{abstract}

\pacs{89.75.Da,02.50.-r}

\date{\today}

\maketitle

\bibliographystyle{apsrev}

Several problems of practical interest require us to understand human
activity patterns \cite{greene97,reynolds03,phone-design}. Typical
examples are the design of telephone systems or web servers, where it is
critical to know how many users would use the service simultaneously.  
The traditional approach to characterize the timing of human activities is
based in two assumptions: the execution of each task is independent from
the others and each task is executed at a constant rate
\cite{fellerII,greene97,reynolds03,phone-design}. A specific task, such as
sending emails or making phone calls, is then modeled as a Poisson process
\cite{fellerII}, characterized by a homogeneous activity pattern. More
precisely the time interval between two consecutive executions of a task
follows an exponential distribution. An increasing amount of empirical
evidence is indicating, however, that human activity patterns are rather
heterogeneous, with short periods of high activity separated by long 
periods of inactivity
\cite{phone-design,Instant,supercomputers,ftp,economic2,barabasi05,dezso05,group}.  
This heterogeneity is characterized by a heavy tail in the distribution of
the time interval between two consecutive executions of the given task
\cite{barabasi05,dezso05,group}.

In practice the execution of one task is not independent for the others.  
Humans keep track of a list of active tasks from where they decide what to
do next, the selection of one task implying the exclusion of the others.
This picture lead Barab\'asi to model the task management by a human as a
queueing system, where the human plays role of the server
\cite{barabasi05}. Queueing systems \cite{gross98} have already received
some attention in the physics literature
\cite{sugiyama:7749,ohira:193,arenas:3196,sole01}. This interest is
motivated by the observation of a non-equilibrium phase transition from a
non-congested phase with a stationary number of active tasks to a
congested phase where the number of active tasks grows in time. In the
non-congested phase the mean waiting time before the execution of an
active task is finite. When approaching the phase transition point the
mean waiting time diverges, while it grows with time in the congested
phase.

The Barab\'asi model belongs, however, to a new class of queueing models
with a fixed number of active tasks. In this case the behavior of interest
comes from the task selection protocol. In the extremal dynamics limit,
when the highest priority task is selected first, numerical simulations
and heuristic arguments show that most of the tasks are executed in one
step, while the waiting time distribution of tasks waiting more than one
step exhibits a heavy tail \cite{barabasi05}. Yet, further research is
required to obtain the scaling behavior in the vicinity of this singular
point.

In this work we obtain exact results for the Barab\'asi model, allowing us
to prove previous conjectures based on heuristic arguments and numerical
simulations, and creating a solid background for future research. We
calculate the priority and waiting time distribution of those tasks
remaining in the list for the case of two active tasks. We corroborate the
observation of a singular behavior in the limit when the task with the
highest priority is selected first, and derive the corresponding scaling
behavior. We also obtain an upper bound for the average waiting time,
which is independent of the selection protocol. Based on this result we
discuss the asymptotic behaviors of the waiting time distribution.  All
the results presented here were checked by numerical simulations,
providing a perfect match with the theoretical curves.

{\it Barab\'asi model:} The Barab\'asi model is defined as follows. A
human keeps track of a list with $L$ active tasks that he/she must do. A
priority $x\geq0$ is assigned to each active task when it is added to the
list, with a probability density function (pdf) $\rho(x)$. The list is
started at $t=0$ by adding $L$ new tasks two it. At each discrete time
step $t>0$ the task in the list with the highest priority is selected
with probability $p$, and with probability $1-p$ a task is selected at
random. The selected task is executed, removed from the list, and a new
task is added. The control parameter $p$ interpolates between the random
selection protocol at $p=0$ and the highest priority first selection
protocol at $p=1$.

The numerical simulations indicate that the case $L=2$ already exhibits
the relevant features of the model \cite{barabasi05}. Furthermore, if we
focus on a single task, such as sending emails, we can model the active
tasks list as a list with two tasks, one corresponding to sending emails
and the other to doing something else. Within this scenario the waiting
time coincides with the time between two consecutive executions of the
corresponding task. Thus, the $L=2$ case provides us with a minimal model
to study the statistical properties of the time between the consecutive
execution of specific tasks.

Consider the Barab\'asi model with $L=2$. The task that has been just
selected and its priority has been reassigned will be called the new task,
while the other task will be called the old task. Let $\rho(x)$ and
$R(x)=\int_0^xdx\rho(x)$ be the priority pdf and distribution
function of the new task, which are given. In turn, let
$\rho_1(x,t)$ and $R_1(x,t)=\int_0^xdx\rho_1(x,t)$ be
the priority pdf and distribution function of the old task at the $t$-th
step. At the $(t+1)$-th step, there are two tasks on the list, their
priorities being distributed according to $R(x)$ and $R_1(x,t)$,
respectively. After selecting one task the old task will have the
distribution function

\begin{equation}
R_1(x,t+1) = \int_0^x dx^\prime \rho_1(x^\prime,t) q(x^\prime)
+ \int_0^{x} dx^\prime \rho(x) q_1(x^\prime,t)\ ,
\label{RtRt}
\end{equation}

\noindent where

\begin{equation}
q(x) = p[1-R(x)] +(1-p)\frac{1}{2}
\label{q}
\end{equation}

\noindent is the probability that the new task is selected given
the old task has priority $x$, and

\begin{equation}
q_1(x,t) = p[1-R_1(x,t)] +(1-p)\frac{1}{2}
\label{q1}
\end{equation}

\noindent is the probability that the old task is selected given the new
task has priority $x$. In the stationary state,
$R_1(x,t+1)=R_1(x,t)$, from (\ref{RtRt}) we obtain

\begin{equation}
R_1(x) = \frac{1+p}{2p} \left[ 1 -
\frac{1}{1+\frac{2p}{1-p}R(x)} \right]\ .
\label{Rx}
\end{equation}

\begin{figure}
\includegraphics[width=3.1in,height=2.0in]{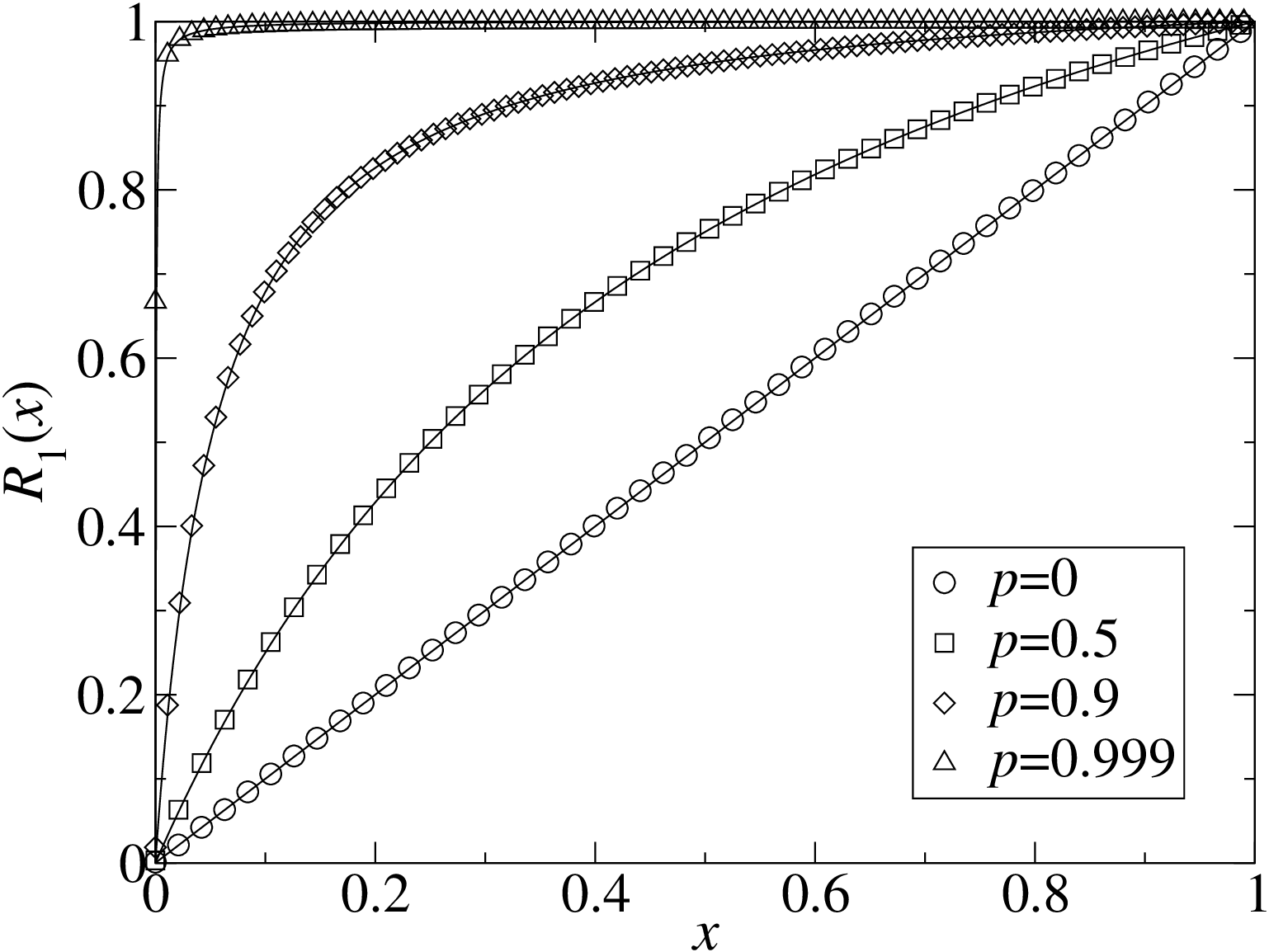}

\caption{Old task priority distribution for the case of a uniform new task
priority distribution function, $\rho(x)=1$ and $R(x)=x$ in $0\leq x\leq
1$, as obtained from (\ref{Rx}) (lines) and numerical simulations
(points). The case $p=0$ corresponds with the random selection protocol
with $R_1(x)=R(x)=x$.}

\label{fig1} 
\end{figure}

To analyze this result let us consider its limiting cases. When
$p\rightarrow0$ (\ref{Rx}) results in

\begin{equation}
\lim_{p\rightarrow0} R_1(x) = R(x)\ .
\label{Rxp0}
\end{equation} 

\noindent Indeed, this limit corresponds to the random selection protocol
and, therefore, the priority distribution of old tasks is equal to that of
new tasks. On the other hand, when $p\rightarrow1$ from (\ref{Rx}) we
obtain

\begin{equation}
\lim_{p\rightarrow1} R_1(x) = \left\{
\begin{array}{ll}
0\ , & x=0\\
1\ , & x>0\ .
\end{array}
\right.
\label{Rxp1}
\end{equation}

\noindent {\it i.e.} $\rho_1(x)$ is concentrated around $x=0$. This result
implies that in the limit $p\rightarrow1$ the new task will always be
selected for execution, resulting in a waiting time $\tau=1$. We are going
to return to this result after computing the waiting time distribution.
The progression between these two limiting cases is illustrated in Fig.
\ref{fig1}, where we plot $R_1(x)$ (\ref{Rx}) as a function of $x$ for a
uniform distribution in $0\leq x\leq1$. As $p$ increases from zero
$R_1(x)$ deviates more from $R(x)$, resulting in a higher probability to
obtain smaller values of $x$. When $p=0.999$, $R_1(x)$ grows to a value of
almost one in a very short $x$-range close to $x=0$, approaching the limit
distribution (\ref{Rxp1}).

Next we turn our attention to the waiting time distribution.  Consider a
task with priority $x$ that has just been added to the queue. The
selection of this task is independent from one step two the other.
Therefore, the probability that it waits $\tau$ steps is given by the
product of the probability that it is not selected in the first $\tau-1$
steps and that it is selected in the $\tau$-th step.  The probability that
it is not selected in the first step is $q_1(x)$, while the probability
that it is not selected in the subsequent steps is $q(x)$. The integration
over the new task's priorities results in

\begin{equation}
P_\tau = \left\{
\begin{array}{ll}
\int_0^\infty dR(x) \left[ 1 - q_1(x) \right] \ , & \tau=1\\
\\
\int_0^\infty dR(x) q_1(x) \left[ 1 - q(x) \right]
q(x)^{\tau-2}\ , & \tau>1\\
\end{array}
\right.
\label{ptau0}
\end{equation}

\begin{figure}
\includegraphics[width=3.1in,height=2.0in]{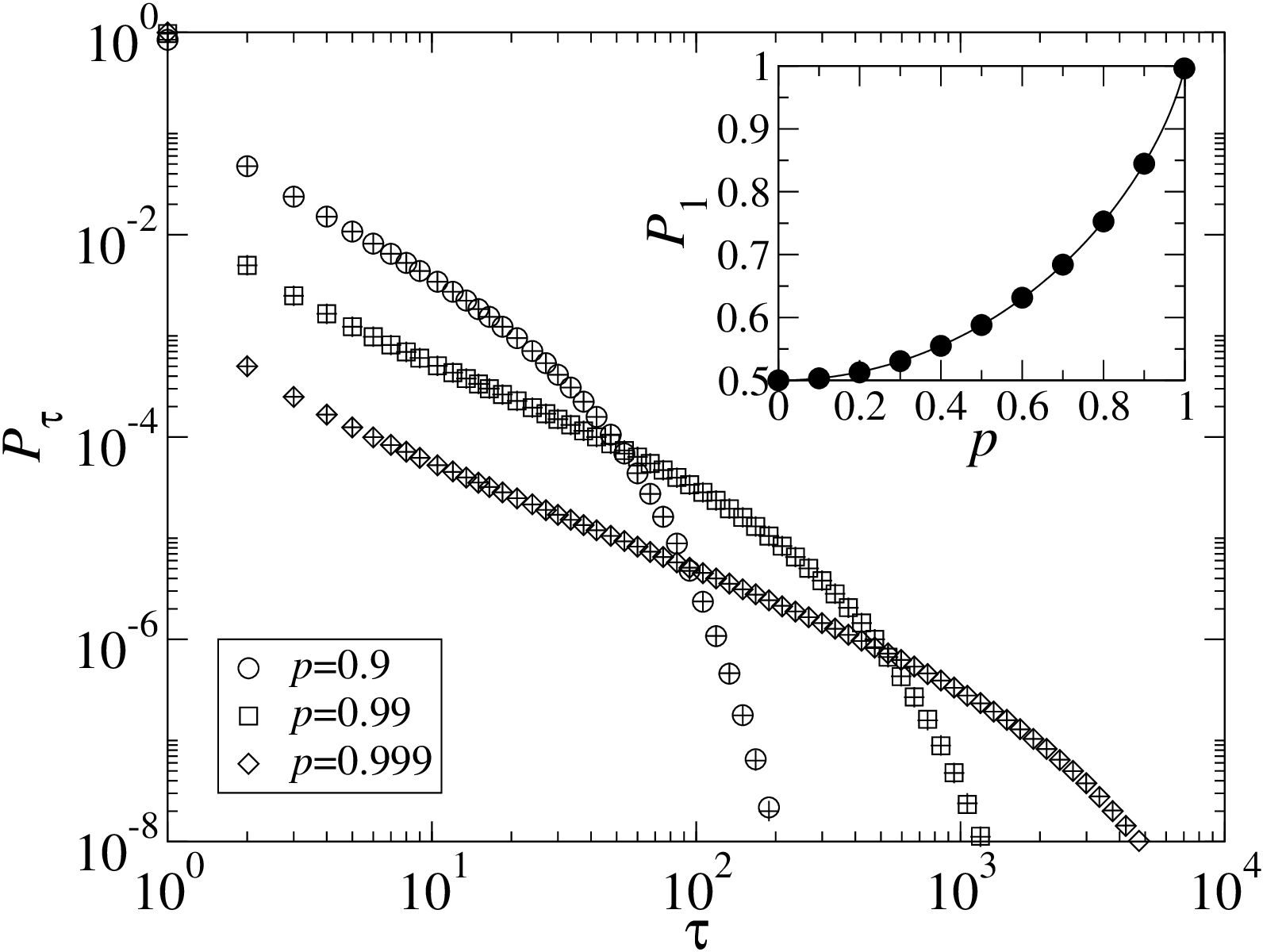}

\caption{Waiting time probability distribution for the case of a uniform
new task priority distribution, $\rho(x)=1$ and $R(x)=x$ in $0\leq x\leq
1$, as obtained from (\ref{ptau}) (pluses) and numerical simulations (open
symbols). The inset shows the fraction of tasks with waiting time $\tau=1$
as a function of $p$, as obtained from (\ref{ptau})  (line) and numerical
simulations (points).}

\label{fig2}
\end{figure}

\noindent Using (\ref{q})-(\ref{Rx}) and integrating (\ref{ptau0}) we 
finally obtain

\begin{equation}
P_\tau = \left\{
\begin{array}{ll}
1 - \frac{1-p^2}{4p} \ln \frac{1+p}{1-p}\ , & \tau=1\\
\\
\frac{1-p^2}{4p} \left[ \left(\frac{1+p}{2}\right)^{\tau-1}
- \left(\frac{1-p}{2}\right)^{\tau-1} \right] \frac{1}{\tau-1}\ , & \tau>1
\end{array}
\right.
\label{ptau}
\end{equation}

\noindent Note that $P_\tau$ is independent of the $\rho(x)$. Indeed, what
matters for a task selection is its relative priority with respect to
other tasks, resulting that all dependences with $x$ in
(\ref{q})-(\ref{Rx}) and (\ref{ptau0}) appears via $R(x)$.

As before, let us consider the limiting cases. In the limit
$p\rightarrow0$ from (\ref{ptau}) it follows that

\begin{equation}
\lim_{p\rightarrow0}P_\tau = \left(\frac{1}{2}\right)^\tau\ ,
\label{ptaup0}
\end{equation}

\noindent for $\tau\geq1$. This limit corresponds with the random
selection protocol, where a task is selected with probability $1/2$ on
each step.  In the other limit, $p\rightarrow1$, we obtain

\begin{equation}
\lim_{p\rightarrow1}P_\tau = \left\{
\begin{array}{ll}
1 + {\cal O}\left(\frac{1-p}{2}\ln(1-p)\right)\ , & \tau=1\\
\\
{\cal O}\left(\frac{1-p}{2}\right) \frac{1}{\tau-1}\ , & \tau>1\ .
\end{array}
\right.
\label{ptaup1}
\end{equation}

\noindent In this case almost all tasks have a waiting time $\tau=1$,
while the waiting time of tasks that are not selected in the first step
follows a power law probability distribution. This picture is illustrated
by a direct plot of $P_\tau$ in (\ref{ptau}). In Fig. \ref{fig2} we plot
$P_\tau$ vs $\tau$ for a uniform distribution in $0\leq x\leq1$. For
$p=0.9$ the probability distribution $P_\tau$ is dominated by an
exponential cutoff. This exponential cutoff can be derived from
(\ref{ptau}) by taking the limit $\tau\rightarrow\infty$ with $p$ fixed,
resulting in

\begin{equation}
P_\tau \sim \frac{1-p^2}{4}\frac{1}{\tau} 
\exp\left( - \frac{\tau}{\tau_0}  \right)\ ,
\label{ptau00}
\end{equation}

\noindent where

\begin{equation}
\tau_0 = \left( \ln \frac{2}{1+p} \right)^{-1}\ .
\label{tau0}
\end{equation}

\noindent When $p\rightarrow1$ we obtain that $\tau_0\rightarrow\infty$
and, therefore, the exponential cutoff is shifted to higher $\tau$ values,
while the power law behavior $P_\tau\sim 1/\tau$ becomes more evident.  
The $P_\tau$ vs $\tau$ curve systematically shifts, however, to lower
values for $\tau>1$, indicating that this power law applies to a vanishing
task fraction (see Fig. \ref{fig2} and (\ref{ptau00})). In turn,
$P_1\rightarrow1$ when $p\rightarrow1$, as it is corroborated by the
direct plot of $P_1$ as a function of $p$ (see inset of Fig. \ref{fig2}).

Another characteristic magnitude of a queueing system is the average
waiting time of an active task before its execution. For $L=2$ we can
calculate the average waiting time directly from (\ref{ptau}), obtaining

\begin{equation}
\langle \tau\rangle = \left\{
\begin{array}{ll}
2 \ , & 0\leq p<1\\
1\ , & p=1\ .
\end{array}
\right.
\label{tauaveL2}
\end{equation}

\noindent This average is restricted to those tasks that are executed and,
therefore, for $p=1$ we are excluding the task that remains indefinitely
in the queue. As we show next, we can extend this result for $L>2$ using
simple arguments.

\begin{figure}
\includegraphics[width=3in,height=2.0in]{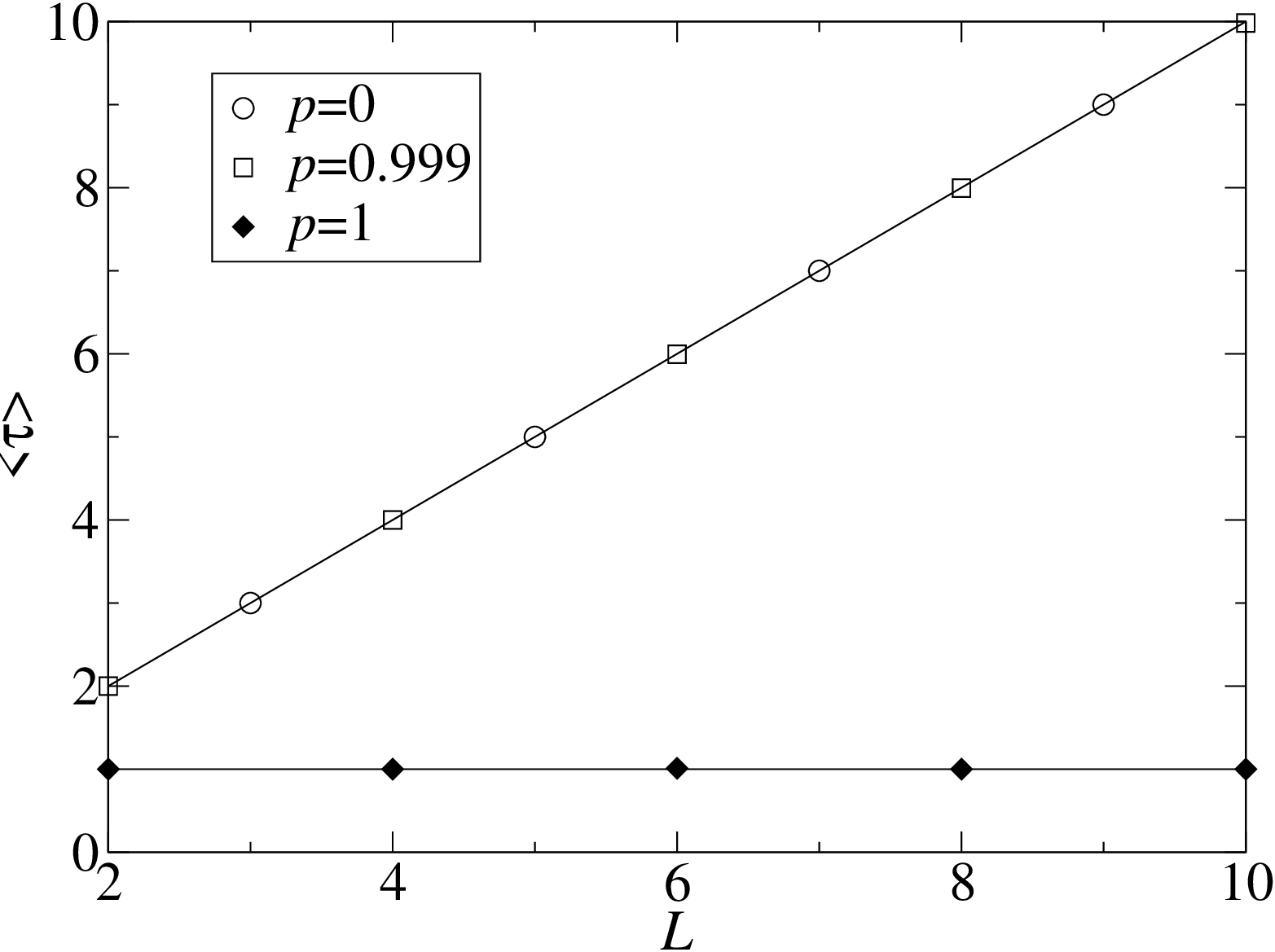}

\caption{Average waiting time of executed tasks vs the list size for the
case of a uniform new task priority distribution function, $\rho(x)=1$ and
$R(x)=x$ in $0\leq x\leq 1$, as obtained from (\ref{tauaveL}) (lines) and
numerical simulations (points).}

\label{fig3} 
\end{figure}

On each step there are $L$ task in the queue and one of them is executed.
Therefore

\begin{equation}
\sum_{i=1}^t\tau_i + \sum_{i=1}^{L-1}\tau_i^\prime = Lt\ ,
\label{tauLt}
\end{equation}

\noindent where $\tau_i$ is the waiting time of the task executed at the
$i$-th step and $\tau_i^\prime$, $i=1,\ldots,L-1$, are the resident times
of the $L-1$ tasks that are still active at the $t$ step. From
(\ref{tauLt}) it follows that

\begin{equation}
\langle \tau\rangle = \lim_{t\rightarrow\infty}
\frac{1}{t}\sum_{i=1}^t\tau_i = L - 
\lim_{t\rightarrow\infty}\frac{1}{t}\sum_{i=1}^{L-1}\tau_i^\prime\ .
\label{tauave}
\end{equation}

\noindent For $0\leq p<1$ the numerical simulations indicate that all
active tasks are expected to be executed \cite{barabasi05}. Therefore
$\langle\tau^\prime\rangle\leq\langle \tau\rangle$ and the last term in
(\ref{tauave}) vanishes when $t\rightarrow\infty$. In contrast, for $p=1$
the numerical simulations \cite{barabasi05} indicate that after some
transient time the most recently added task is always executed, while
$L-1$ tasks remain indefinitely in the queue. In this case
$\tau_i^\prime\sim t$ when $t\rightarrow\infty$ and the last term in
(\ref{tauave}) is of the order of $L-1$ when $t\rightarrow\infty$.  Based
on these arguments we conjecture that the average waiting time of executed
tasks is given by

\begin{equation}
\langle \tau\rangle = \left\{
\begin{array}{ll}
L \ , & 0\leq p<1\\
1\ , & p=1\ .
\end{array}
\right.
\label{tauaveL}
\end{equation}

\noindent This results was proved for $L=2$ (\ref{tauaveL2}), and for
$L>2$ it is corroborated by the numerical simulations (see Fig.
\ref{fig3}). It is important to note that the equality in (\ref{tauave})
is independent of the selection protocol, allowing us to reach more
general conclusions beyond the Barab\'asi model.  Using (\ref{tauave}) we
obtain

\begin{equation}
\langle\tau\rangle \leq L\ .
\label{tauaveLineq}
\end{equation}

\noindent From this constraint it follows that $P_\tau$ must decay faster
than $\tau^{-2}$ when $\tau\rightarrow\infty$. Thus, when
$\tau\rightarrow\infty$ either

\begin{equation}
P_\tau \sim a\tau^{-\alpha}\ ,
\label{ptau1}
\end{equation}

\noindent where $a$ is a constant and $\alpha>2$, or

\begin{equation}
P_\tau = \tau^{-\alpha} f \left( \frac{\tau}{\tau_0} \right)\ ,
\label{ptauscaling}
\end{equation}

\noindent where $\tau_0>0$ and $f(x)={\cal O}(bx^{\alpha-2})$ when
$x\rightarrow\infty$, where $b$ is a constant. For instance, for the
Barab\'asi model with $L=2$ and $0\leq p<1$, $\alpha=1$ and $f(x)$ decays
exponentially (\ref{ptau00}), in agreement with (\ref{ptauscaling}).

The empirical evidence \cite{barabasi05,dezso05,group} is in favor of the
second scenario with $\alpha=1$. This observation is in agreement with our
expectation of a natural cutoff. For instance, we might go on a trip and
not check emails for several days, but sooner or later we are going to do
it because we receive and transmit important information using this
communication media. This cutoff time is expected to be more related to
the necessity of performing a given task rather than to the interaction
with other tasks. In this sense, the random selection of a task in the
Barab\'asi model could be interpreted as a task whose priority suddenly
increases to the maximum value because the need to execute it after being
on the queue for a long time. This indicates future directions of
research, considering the case when the priority of old tasks may also
change with time \cite{blanchard05}.

The singular behavior of the Barab\'asi model is a consequence of the
extremal dynamics rule: the task with the highest priority is selected
first. Therefore, the conclusions obtained here are also relevant to other
complex system evolving with extremal dynamics
\cite{bak93,slanina99,boettcher01}. In this more general context the
waiting time represents the life time before selection, an important
quantity in evolution models \cite{bak93,slanina99} and optimization
methods based in extremal dynamics \cite{boettcher01}. Further research is
required, however, to determine the influence of other factors such as
correlations among neighbors, which are absent in the Barab\'asi queueing
model.

This work was supported by NSF ITR 0426737, NSF ACT/SGER 0441089 awards
and the James S. McDonnell Foundation.


\begin{thebibliography}{17}
\expandafter\ifx\csname natexlab\endcsname\relax\def\natexlab#1{#1}\fi
\expandafter\ifx\csname bibnamefont\endcsname\relax
  \def\bibnamefont#1{#1}\fi
\expandafter\ifx\csname bibfnamefont\endcsname\relax
  \def\bibfnamefont#1{#1}\fi
\expandafter\ifx\csname citenamefont\endcsname\relax
  \def\citenamefont#1{#1}\fi
\expandafter\ifx\csname url\endcsname\relax
  \def\url#1{\texttt{#1}}\fi
\expandafter\ifx\csname urlprefix\endcsname\relax\def\urlprefix{URL }\fi
\providecommand{\bibinfo}[2]{#2}
\providecommand{\eprint}[2][]{\url{#2}}

\bibitem[{\citenamefont{Anderson}(2003)}]{phone-design}
\bibinfo{author}{\bibfnamefont{H.~R.} \bibnamefont{Anderson}},
  \emph{\bibinfo{title}{Fixed Broadband Wireless System Design}}
  (\bibinfo{publisher}{Wiley, New York}, \bibinfo{year}{2003}).

\bibitem[{\citenamefont{Greene}(1997)}]{greene97}
\bibinfo{author}{\bibfnamefont{J.~H.} \bibnamefont{Greene}},
  \emph{\bibinfo{title}{Production and inventory control handbook}}
  (\bibinfo{publisher}{MacGraw-Hill, New York}, \bibinfo{year}{1997}).

\bibitem[{\citenamefont{Reynolds}(2003)}]{reynolds03}
\bibinfo{author}{\bibfnamefont{P.}~\bibnamefont{Reynolds}},
  \emph{\bibinfo{title}{Call center staffing}} (\bibinfo{publisher}{The Call
  Center School Press, Lebanon, Tennessee}, \bibinfo{year}{2003}).

\bibitem[{\citenamefont{Feller}(1966)}]{fellerII}
\bibinfo{author}{\bibfnamefont{W.}~\bibnamefont{Feller}},
  \emph{\bibinfo{title}{An introduction to probability theory and its
  applications}} (\bibinfo{publisher}{Wiley, New York}, \bibinfo{year}{1966}),
  \bibinfo{note}{vol. II}.

\bibitem[{\citenamefont{Barab\'asi}(2005)}]{barabasi05}
\bibinfo{author}{\bibfnamefont{A.-L.} \bibnamefont{Barab\'asi}},
  \bibinfo{journal}{Nature} \textbf{\bibinfo{volume}{435}},
  \bibinfo{pages}{207} (\bibinfo{year}{2005}).

\bibitem[{\citenamefont{Masoliver et~al.}(2003)\citenamefont{Masoliver,
  Montero, and Weiss}}]{economic2}
\bibinfo{author}{\bibfnamefont{J.}~\bibnamefont{Masoliver}},
  \bibinfo{author}{\bibfnamefont{M.}~\bibnamefont{Montero}}, \bibnamefont{and}
  \bibinfo{author}{\bibfnamefont{G.~H.} \bibnamefont{Weiss}},
  \bibinfo{journal}{Phys. Rev. E} \textbf{\bibinfo{volume}{67}},
  \bibinfo{pages}{021112} (\bibinfo{year}{2003}).

\bibitem[{\citenamefont{Dewes et~al.}(2003)\citenamefont{Dewes, Wichmann, and
  Feldman}}]{Instant}
\bibinfo{author}{\bibfnamefont{C.}~\bibnamefont{Dewes}},
  \bibinfo{author}{\bibfnamefont{A.}~\bibnamefont{Wichmann}}, \bibnamefont{and}
  \bibinfo{author}{\bibfnamefont{A.}~\bibnamefont{Feldman}}, in
  \emph{\bibinfo{booktitle}{Proc. 2003 ACM SIGCOMM Conf. on Internet
  Measurement (IMC-03)}} (\bibinfo{publisher}{ACM Press, New York},
  \bibinfo{year}{2003}).

\bibitem[{\citenamefont{Kleban and Clearwater}(2003)}]{supercomputers}
\bibinfo{author}{\bibfnamefont{S.~D.} \bibnamefont{Kleban}} \bibnamefont{and}
  \bibinfo{author}{\bibfnamefont{S.~H.} \bibnamefont{Clearwater}}, in
  \emph{\bibinfo{booktitle}{Proc. of SC'03, November 15-21, Phonenix, AZ, USA}}
  (\bibinfo{year}{2003}).

\bibitem[{\citenamefont{Paxson and Floyd}(1995)}]{ftp}
\bibinfo{author}{\bibfnamefont{V.}~\bibnamefont{Paxson}} \bibnamefont{and}
  \bibinfo{author}{\bibfnamefont{S.}~\bibnamefont{Floyd}},
  \bibinfo{journal}{IEEE/ACM Transactions in Networking}
  \textbf{\bibinfo{volume}{3}}, \bibinfo{pages}{226} (\bibinfo{year}{1995}).


\bibitem[{\citenamefont{Barab\'asi et~al.}()\citenamefont{Barab\'asi,
  Dezs\H{o}, Oliveira, and V\'azquez}}]{group}
  \bibinfo{author}{\bibfnamefont{A.}~\bibnamefont{V\'azquez}},
  \bibinfo{author}{\bibfnamefont{Z.}~\bibnamefont{Dezs\H{o}}},
  \bibinfo{author}{\bibfnamefont{J.}~\bibnamefont{Oliveira}},
  \bibinfo{author}{\bibfnamefont{K.-I.}~\bibnamefont{Goh}},
  \bibinfo{author}{\bibfnamefont{I.}~\bibnamefont{Kondor}}, 
\bibnamefont{and}
  \bibinfo{author}{\bibfnamefont{A.-L.} \bibnamefont{Barab\'asi}},
  \bibinfo{note}{arXive:physics/0510117}.

\bibitem[{\citenamefont{Dezs\H{o} et~al.}()\citenamefont{Dezs\H{o}, Almaas,
  Luk\'acs, and B.~R\'acz}}]{dezso05}
\bibinfo{author}{\bibfnamefont{Z.}~\bibnamefont{Dezs\H{o}}},
  \bibinfo{author}{\bibfnamefont{E.}~\bibnamefont{Almaas}},
  \bibinfo{author}{\bibfnamefont{A.}~\bibnamefont{Luk\'acs}}, \bibnamefont{and}
  \bibinfo{author}{\bibfnamefont{A.-L.~B.} \bibnamefont{B.~R\'acz},
  \bibfnamefont{I.~Szakad\'at}}, \bibinfo{note}{arXive:physics/0505087}.

\bibitem[{\citenamefont{Gross and Harris}(1998)}]{gross98}
\bibinfo{author}{\bibfnamefont{D.}~\bibnamefont{Gross}} \bibnamefont{and}
  \bibinfo{author}{\bibfnamefont{C.~M.} \bibnamefont{Harris}},
  \emph{\bibinfo{title}{Queueing theory}} (\bibinfo{publisher}{John Wiley \&
  Sons, New York}, \bibinfo{year}{1998}).

\bibitem[{\citenamefont{Sugiyama and Yamada}(1997)}]{sugiyama:7749}
\bibinfo{author}{\bibfnamefont{Y.}~\bibnamefont{Sugiyama}} \bibnamefont{and}
  \bibinfo{author}{\bibfnamefont{H.}~\bibnamefont{Yamada}},
  \bibinfo{journal}{Phys. Rev. E} \textbf{\bibinfo{volume}{55}},
  \bibinfo{pages}{7749} (\bibinfo{year}{1997}).

\bibitem[{\citenamefont{Ohira and Sawatari}(1998)}]{ohira:193}
\bibinfo{author}{\bibfnamefont{T.}~\bibnamefont{Ohira}} \bibnamefont{and}
  \bibinfo{author}{\bibfnamefont{R.}~\bibnamefont{Sawatari}},
  \bibinfo{journal}{Phys. Rev. E} \textbf{\bibinfo{volume}{58}},
  \bibinfo{pages}{193} (\bibinfo{year}{1998}).

\bibitem[{\citenamefont{Arenas et~al.}(2001)\citenamefont{Arenas, Diaz-Guilera,
  and Guimera}}]{arenas:3196}
\bibinfo{author}{\bibfnamefont{A.}~\bibnamefont{Arenas}},
  \bibinfo{author}{\bibfnamefont{A.}~\bibnamefont{Diaz-Guilera}},
  \bibnamefont{and} \bibinfo{author}{\bibfnamefont{R.}~\bibnamefont{Guimera}},
  \bibinfo{journal}{Phys. Rev. Lett.} \textbf{\bibinfo{volume}{86}},
  \bibinfo{pages}{3196} (\bibinfo{year}{2001}).

\bibitem[{\citenamefont{Sol\'e and Valverde}(2001)}]{sole01}
\bibinfo{author}{\bibfnamefont{R.}~\bibnamefont{Sol\'e}} \bibnamefont{and}
  \bibinfo{author}{\bibfnamefont{S.}~\bibnamefont{Valverde}},
  \bibinfo{journal}{Physica A} \textbf{\bibinfo{volume}{289}},
  \bibinfo{pages}{595} (\bibinfo{year}{2001}).

\bibitem[{\citenamefont{Blanchard and Hongler}()}]{blanchard05}
\bibinfo{author}{\bibfnamefont{P.}~\bibnamefont{Blanchard}} \bibnamefont{and}
  \bibinfo{author}{\bibfnamefont{M.-O.} \bibnamefont{Hongler}},
  \bibinfo{note}{preprint}.

\bibitem{bak93} P. Bak and K. Sneppen, Phys. Rev. Lett. {\bf 71}, 4093 
(1993).

\bibitem{slanina99} F. Slanina and M. Kotrla, Phys. Rev. Lett. {\bf 83}, 
5587 (1999).

\bibitem{boettcher01} S. Boettcher and A. G. Percus, Phys. Rev. Lett. {\bf 
86}, 5211 (2001).

\end{thebibliography}
\end{document}